\documentclass[fleqn,10pt]{wlscirep}
\providecommand{\e}[1]{\ensuremath{\times 10^{#1}}}

\title{Thermally and field-driven mobility of emergent magnetic charges in square artificial spin ice}

\author[1,2*]{Sophie~A.~Morley}
\author[3,4]{Jose Maria~Porro}
\author[1]{Ale\v{s}~Hrabec}
\author[5]{Mark~C.~Rosamond}
\author[5]{Edmund~H.~Linfield}
\author[1]{Gavin Burnell}
\author[6,7]{Mi-Young~Im}
\author[2,8]{Peter~J.~Fischer}
\author[4]{Sean~Langridge}
\author[1,$\dagger$]{Christopher~H.~Marrows}

\affil[1]{School of Physics and Astronomy, University of Leeds, Leeds LS2 9JT, United Kingdom}
\affil[2]{Department of Physics, University of California, Santa Cruz, California, 95064, USA}
\affil[3]{ISIS Neutron and Muon Source, STFC Rutherford Appleton Laboratory, Chilton, Didcot, Oxon. OX11 0QX, United Kingdom}
\affil[4]{BCMaterials, Basque Center for Materials, Applications and Nanostructures, 48940 Leioa, Spain}
\affil[5]{School of Electronic and Electrical Engineering, University of Leeds, Leeds LS2 9JT, United Kingdom}
\affil[6]{Center for X-ray Optics, Lawrence Berkeley National Laboratory, 1 Cyclotron Road, Berkeley, CA 94720, USA}
\affil[7]{Daegu Gyeongbuk Institute of Science and Technology, Daegu 711-873, Korea}
\affil[8]{Materials Sciences Division, Lawrence Berkeley National Laboratory, 1 Cyclotron Road, Berkeley, CA 94720, USA}
\affil[*]{samorley@ucsc.edu}
\affil[$\dagger$]{c.h.marrows@leeds.ac.uk}

\begin{abstract}
Designing and constructing model systems that embody the statistical mechanics of frustration is now possible using nanotechnology. We have arranged nanomagnets on a two-dimensional square lattice to form an artificial spin ice, and studied its fractional excitations, emergent magnetic monopoles, and how they respond to a driving field using X-ray magnetic microscopy. We observe a regime in which the monopole drift velocity is linear in field above a critical field for the onset of motion. The temperature dependence of the critical field can be described by introducing an interaction term into the Bean-Livingston model of field-assisted barrier hopping. By analogy with electrical charge drift motion, we define and measure a monopole mobility that is larger both for higher temperatures and stronger interactions between nanomagnets. The mobility in this linear regime is described by a creep model of zero-dimensional charges moving within a network of quasi-one-dimensional objects. 
\end{abstract}
\begin{document}

\flushbottom
\maketitle

\thispagestyle{empty}


Artificial spin ices (ASI) are arrays of nanomagnetic islands that are effectively single domain and so have bistable Ising-like macrospin states. They have emerged as a playground to study the statistical mechanics of frustration phenomena in a simplified setting \cite{Nisoli2013}. The particular beauty of these systems is that one is able to continuously tune various parameters such as interaction strength between nanomagnets \cite{Zhang2013}, material \cite{Drisko2015}, and array topology and geometry \cite{Gilbert2014,Gilbert2016}, making them designer metamaterials. There has been great interest in the past in the attainment and preparation of low energy or ground states and the impact different parameters have on this, imaged as athermal snapshots of quenched thermally equilibrated states \cite{morgan,porro,Zhang2013}. More recently, partly or fully thermalised systems undergoing fluctuation and relaxation \cite{melting,farhansquare,hypercube,Kapaklis2014} and temperature driven phase transitions \cite{Anghinolfi2015} have been studied. Here we study the thermally-activated dynamics these systems display in response to an external drive field through the direct imaging of the microstates involved. 

\begin{figure}[tb]
\centering
\includegraphics[width=16cm]{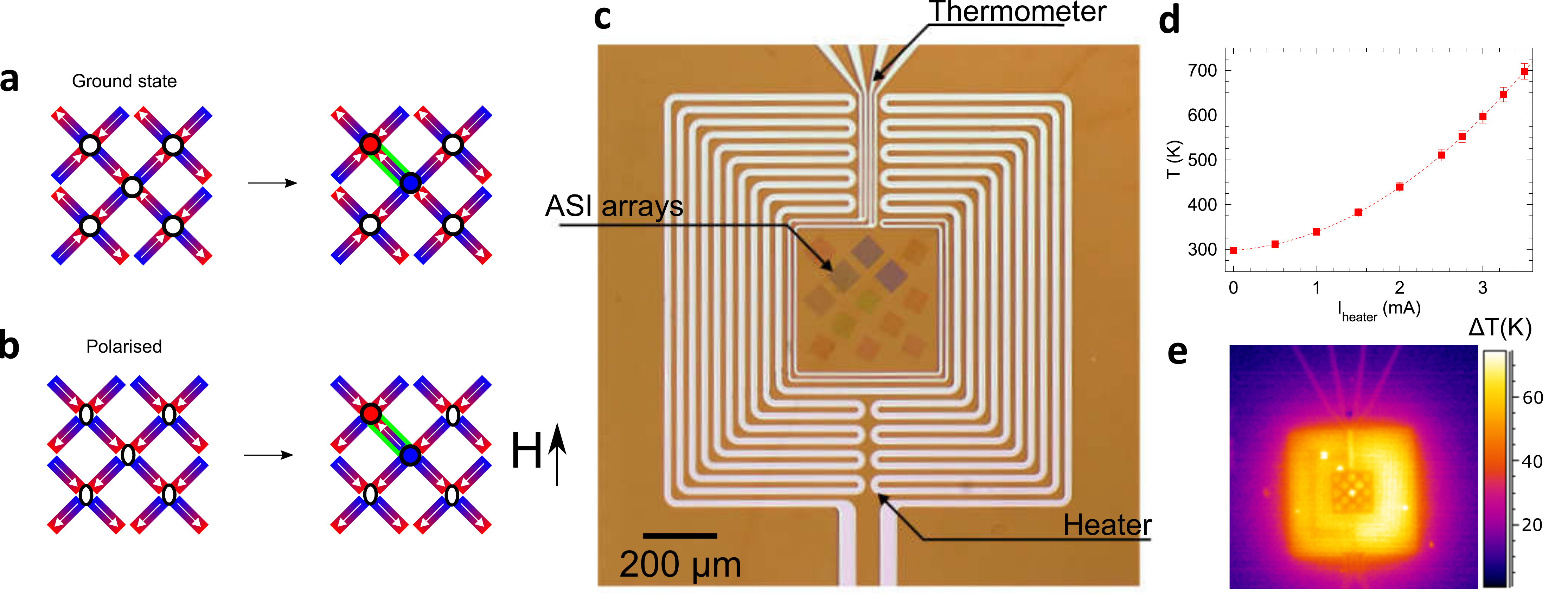}
\caption{\textsf{\textbf{Monopole-antimonopole pair creation and on-membrane heater.} Starting from a lattice of ice-rule obeying vertices (left) it is possible to flip a connecting macrospin between two vertices (outlined in green on the right) to create two Type 3 vertices, known as an emergent monopole-antimonopole pair. This can be done from \textbf{a}, the Type 1 ground state vertices by thermal excitation or \textbf{b}, from a pair of diagonally field-polarised Type 2 vertices by applied field. \textbf{c}, An optical micrograph of the on-membrane heater and thermometer. The meandering Pt wires are counter-wound to avoid induced magnetic field at the sample and the ASI arrays were patterned in the centre of the design: here 13 separate arrays are visible with various lattice constants. \textbf{d}, The temperature of the thermometer, obtained as a function of the heater current, $I_\mathrm{heater}$. \textbf{e}, An infra-red thermal microscope image taken at $I_\mathrm{heater}$~=~2.3~mA ($T$~=~467~K), showing an even temperature distribution across almost all the arrays on the membrane.}}
\label{paircreation}
\end{figure}

In particular, we take advantage of the concept that excitations in the pyrochlore spin ice materials have been described as emergent magnetic monopole-like charges \cite{Castelnovo2008}. Experimentally confirmed in the pyrochlore systems \cite{Morris2009,Fennell2009}, emergent magnetic monopoles have been observed in kagome \cite{ladak,Mengotti2011} and square \cite{Pollard2012} artificial spin ices at the coercive field during athermal reversal. The picture for the square ices is the same as in the pyrochlores: these systems obey an `ice rule', defined as the lowest energy arrangements consisting of two moments pointing into the fourfold vertex and two pointing out, leading to no net charge at the vertex in a dumbbell model. Violations of the rule--excitations--lead to net vertex charges. In the square ASI studied here there are two ice-rule obeying vertex types as depicted in Fig.~\ref{paircreation}a and b.  The first, known as Type 1 ($T_{1}$) in the standard scheme\cite{wang}, has the lowest energy and is the ground state vertex where the islands which are closest together (perpendicular neighbours) have the less energetically-costly interaction of head-to-tail (Fig.~\ref{paircreation}a). In the field-polarised state, known as Type 2 ($T_{2}$), this more favourable interaction of head-to-tail is between the second nearest neighbours (parallel islands), giving a slightly higher overall energy (Fig.~\ref{paircreation}b). The difference in energy is due to the reduced dimensionality of the ASI\cite{moller}, although the degeneracy can be restored by introducing height offsets so that the system is no longer truly two-dimensional \cite{Perrin2016}. Starting from either of these ice-rule obeying states it is possible to flip the macrospin of a nanomagnet in order to change the vertex configurations to Type 3 ($T_{3}$) and consequently violate the ``two-in/two-out" ice rule. As shown on the right-hand side of Fig.~\ref{paircreation}a and b, this results in three-in and one-out state for one vertex (red circle) and three-out one-in for the other (blue circle), which possess opposite net magnetic charges at the vertex and can be thought of as an emergent monopole-antimonopole pair. These may then move apart by flipping further spins, leaving a flux-carrying chain of vertices that is often termed a Dirac string (see Fig.~\ref{imagingmps}), although it is truly of Nambu form in the two-dimensional system\cite{Silva2013}.

The ability to drive these emergent magnetic charges with applied field is analogous to using electric fields to drive a current of electrical charge carriers. The term ‘magnetricity’ was first coined to describe this effect in the pyrochlore spin ice materials \cite{Bramwell2009,Giblin2011}. Here we demonstrate the effect of an applied magnetic field on the movement of magnetic charges in the thermally activated regime within our square ASI systems. Many of the imaging methods that have been suited to probe the dynamics of ASI so far exploit the properties of electrons, such as photoemission electron microscopy \cite{Mengotti2011} and Lorentz transmission electron microscopy \cite{Pollard2012}. Therefore, there has been limited exploration of the thermal fluctuations under a driven magnetic field. Also, due to the high temperatures or low moment materials required for thermal behaviour, magnetic force microscopy (MFM) has not been suitable for dynamic studies of thermally active arrays, since the stray field of the tip would perturb the fluctuating states. The ‘photon-in-photon-out’ nature of the magnetic transmission X-ray microscopy (MTXM) method used here presents a unique opportunity to probe the monopole dynamics as a function of field and temperature. Just as electrons and holes in a semiconductor are driven in opposite directions by an electric field, here we can drive apart the opposite magnetic charges in a pair created from the ice rule state with a magnetic field. By directly observing this motion using MTXM, we have observed increased mobility of these charges with temperature and coupling strength of the islands, similar to the properties of ionic hopping conduction of electrons in solids \cite{Rao1975}. These tend to depend on the probability of hopping between sites with a varying potential energy landscape, akin to that which would be expected in a system with a varying degree of disorder in the coercive fields inherent from the patterning process \cite{PhysRevB.84.180412}. We find a critical field for the onset of motion of the monopole charges, beyond which the thermally activated drift velocity is linear in drive field. This linear creep regime reveals an emergent reduction in the dimensionality of the system. 

\section*{Results}

\subsection*{Magnetic x-ray imaging}

Our experiments were performed on square ASI arrays with two different lattice spacings, $a$, of 350 and 400~nm. The 7~nm thick Co$_{60}$Fe$_{20}$B$_{20}$ islands were nominally 250~nm~$\times$~80~nm in lateral size and were fabricated on soft x-ray-transparent Si$_3$N$_4$ membranes. In order to probe the thermally-activated monopole drift motion, we developed an on-membrane heater and thermometer, as shown in Fig.~\ref{paircreation}c. The lithographically patterned heater can raise the temperature $T$ of the thermometer and the enclosed ASI arrays to values in excess of 700~K, repeatedly. Our islands are large enough that their macrospins are frozen at room temperature, since the energy barrier for reversal $E_\mathrm{b} \gg k_\mathrm{B}T_{\mathrm{room}}$, providing thermally stable states for imaging. Before each image was taken, a current $I_\mathrm{heater}$ was applied for 100~ms to raise the temperature to a value where magnetisation dynamics--under field, if one was applied--can take place. The calibration is shown in Fig.~\ref{paircreation}d and a thermal microscopy image shows an even temperature distribution across the patterned area in Fig.~\ref{paircreation}e. The thermal mass of the membrane is so small that heating and cooling on switching the heater current on and off is effectively instantaneous, freezing the state at the end of the $I_\mathrm{heater}$ pulse ready for imaging. Each image acquired was the average of $8 \times0.8$~s exposures.

\begin{figure}[tb]
\centering
\includegraphics[width=11cm]{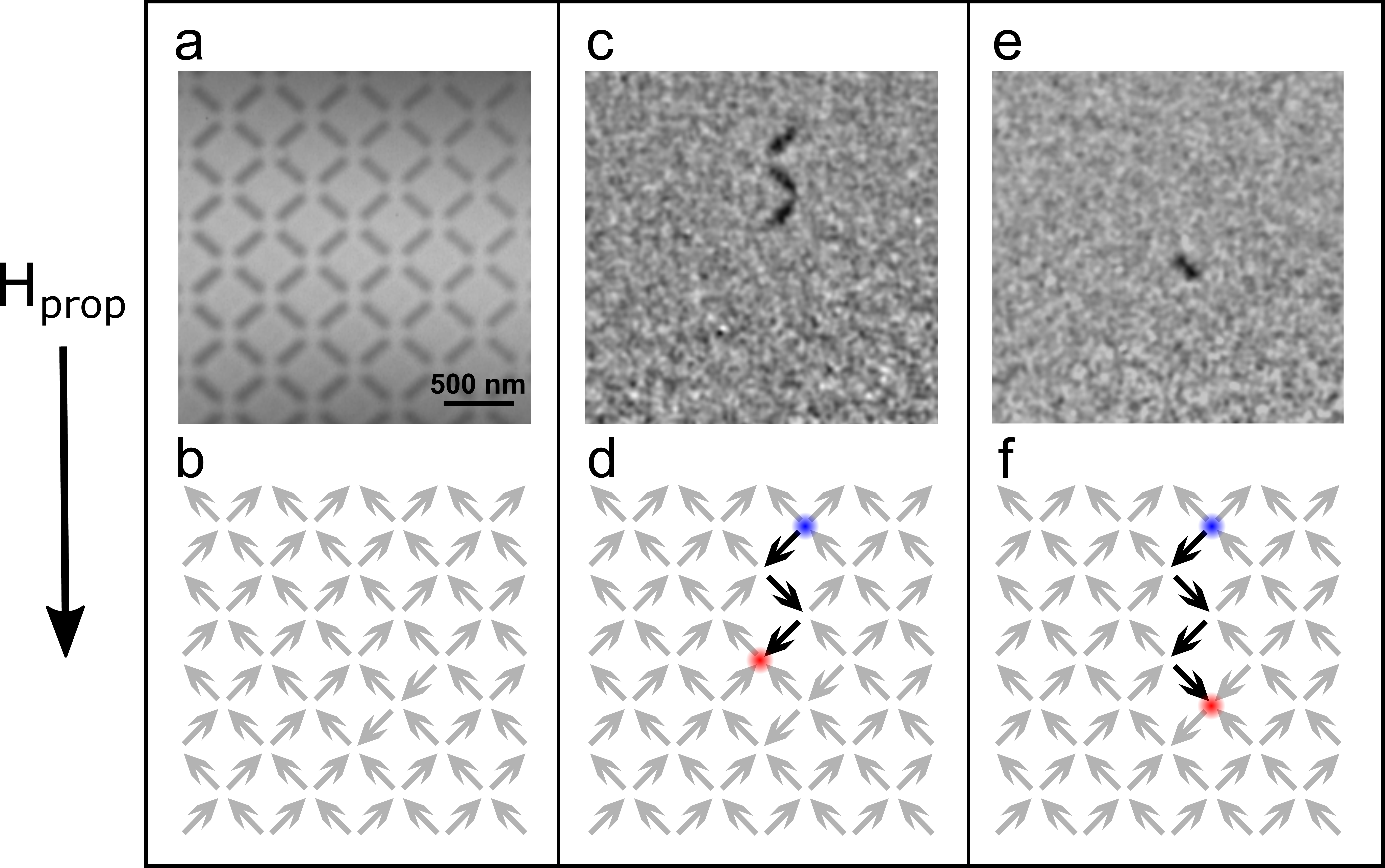}
\caption{\textsf{\textbf{Imaging monopole motion.} \textbf{a}, The raw absorption image for the 400~nm lattice spacing sample and \textbf{b}, a schematic of the initial, pure $T_2$, diagonally field-polarised state. \textbf{c}, A difference image where three islands have reversed and \textbf{d}, the schematic of the corresponding injected monopole-antimonopole ($T_3$) pair indicated with a red and blue circle at the ends of the string of reversed islands ($T_1$ vertices) in black. \textbf{e}, The next consecutive difference image where another island has reversed and \textbf{f}, the cumulative motion of the monopole-antimonopole pair, schematically showing the creation of an additional $T_1$ vertex and growth of the Nambu string.}}
\label{imagingmps}
\end{figure}

\begin{figure*}[tb]
\centering
\includegraphics[width=17cm]{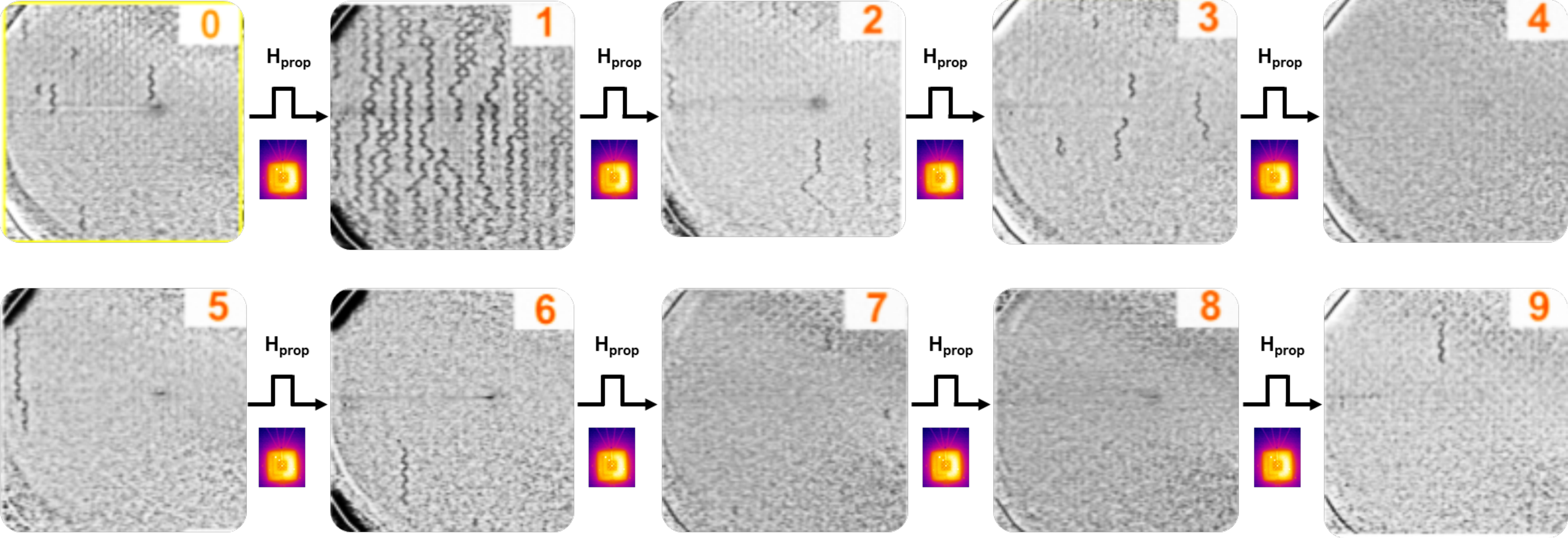}
\caption{\textsf{\textbf{Image sequence for 350~nm lattice spacing sample.} The first image (`0'), marked by a yellow box, is the injection state with a small number of monopole-antimonopole pairs joined by short strings. Each subsequent image is the result of a 100~ms heating pulse to a temperature of 495~K under a 63~mT propagation field. Black contrast indicates islands that have reversed with respect to the previous image.}}
\label{pulses}
\end{figure*}

An example of a raw soft X-ray absorption image taken is shown in Fig.~\ref{imagingmps}a, with the $a = 400$~nm ASI array clearly visible. To perform our experiment, we first applied a large saturating field of -73~mT at a temperature of 439~K in order to create a full diagonally polarised $T_{2}$ background, as shown schematically in Fig.~\ref{imagingmps}b. Then a positive injection field of 64~mT (smaller than the room temperature coercivity) was applied for 100~ms with zero heater current. This created the injection state, which was engineered so that a low number of monopole-antimonopole excitations were created above the $T_{2}$ background, in order to be able to easily follow their trajectories within the field-of-view. The difference image between the saturated state and an injection is shown in Fig.~\ref{imagingmps}c with its corresponding monopole-antimonopole pair of $T_3$ vertices highlighted as red and blue circles in Fig.~\ref{imagingmps}d. They separate via chains of $T_1$ vertices connected by reversed islands, which form Nambu strings\cite{Silva2013}, shown in the figure as the line of black islands tracing out the path of monopole motion. Once driven with a pulse of magnetic field and temperature, further islands will reverse propagating the charge and lengthening the string as demonstrated in Fig.~\ref{imagingmps}e and f. The variable length of these strings means that they can be considered to be a form of avalanche, with longer strings being less common than shorter ones. Hence, an appropriate analogy with thermally activated drift motion of electrical charges in a variable range hopping regime can be drawn. 

A full sequence of difference images is presented in Fig.~\ref{pulses} for the $a = 350$~nm array. Highlighted in the yellow box of the figure is the injection state, which contains five separate chains of reversed islands with lengths between 2-6 islands. Each chain marks the separation path of a monopole-antimonopole pair at its ends. Just before each subsequent image a propagation field $H_{\mathrm{prop}} = 63$~mT and a temperature of 495~K was applied for 100~ms, after which the array was frozen down to room temperature and the resultant string propagation was imaged. This field value is too small to cause any change in the state of the array at room temperature, it is necessary to heat the array in order to induce thermally-activated dynamics under field. It can be seen from the image sequence that there are many initial events after the first propagation field pulse where many strings can be seen within the field of view, after this there are fewer per image but they tend to have string lengths, $L > 4$ islands. This was found to be in contrast to an image sequence with the same propagation field but a lower temperature of 467~K, where there were many fewer reversals in the same measurement time and actually after the fourth pulse, no more islands reversed.

\subsection*{Monopole velocity measurement}

\begin{figure}[tb]
\centering
\includegraphics[width=11cm]{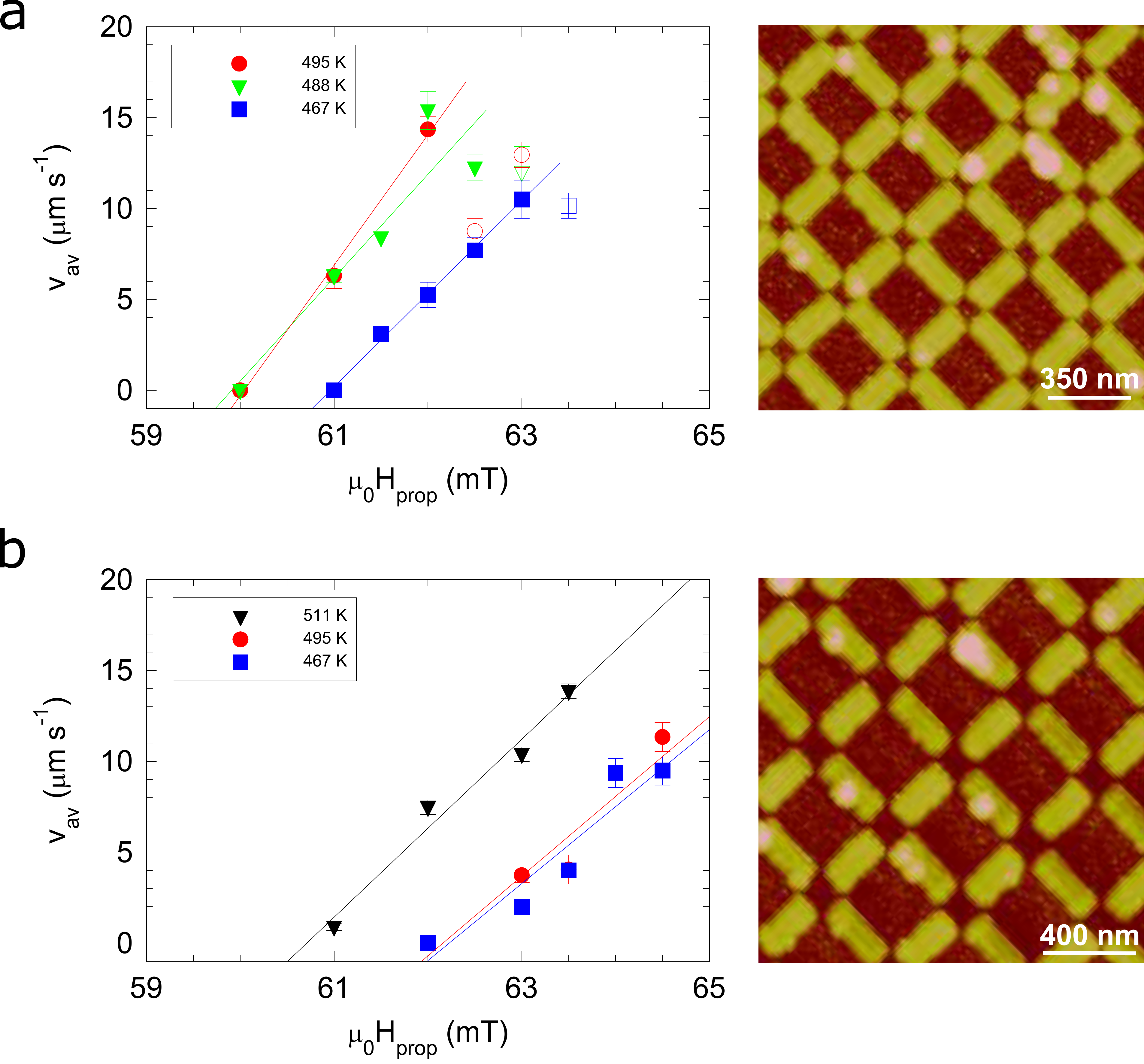}
\caption{\textsf{\textbf{Monopole velocities.} \textbf{a}, The measured velocity of the monopole excitations in the $a = 350$~nm array is plotted as a function of the propagation field. Higher velocities are reached at lower fields for increased temperature.  \textbf{b}, For the more weakly interacting, $a = 400$~nm, sample the velocities are plotted on the same scale for comparison but at a slightly extended temperature range. The solid lines show straight line fits to the data in the linear regime just above the critical field for the onset of motion. Data points at higher fields, in the regime when strings start to overlap (indicated by open symbols), are neglected in the fits. An atomic force micrograph for each sample is also shown on the right of each panel.}}
\label{mpvelocity}
\end{figure}

The average monopole velocity $v_\mathrm{av}$ was measured in each case, which was defined as the average string length per frame over the entire pulse sequence (i.e. a total of 10~pulses for a total duration of 1~s, during which thermally activated dynamics took place). This gave a value of 37~lattice~hops~s$^{-1}$ at $T$~=~495~K, compared to 30~lattice~hops~s$^{-1}$ for $T$~=~467~K. The mean observed string length was also longer for the higher temperature, $10.1 \pm 0.8$~islands compared to $7.8 \pm 0.9$~islands. As previously mentioned, the separation of these oppositely charged emergent monopoles can be likened to the flow of electric charge, and in order to draw a parallel to this physics we measured the average velocity using various propagation fields at different temperatures. The results are shown in Fig.~\ref{mpvelocity}a for the $a = 350$~nm array. The velocity for each temperature has a similar form: there is a critical field below which no monopole motion was observed, after which there is a regime that is a few mT wide where the velocity is linear in field. At higher fields an apparent departure from linearity occurs, due to strings growing so rapidly that they start to overlap, which makes the measurement of individual monopole velocities unreliable. 

We performed an equivalent set of measurements for the less strongly interacting ASI, in which the islands were the same size but the lattice spacing was increased to 400~nm. The most striking difference is the average string length is greatly reduced to $2.2\pm 0.4$~islands, less than a quarter of that observed for the 350~nm lattice spacing for the same temperature and propagation field ($T$ = 495~K and $H_{\mathrm{prop}} = 63$~mT). The average emergent monopole velocity is more than four times as slow: 8.3~lattice~hops~s$^{-1}$ c.f. 37~lattice~hops~s$^{-1}$ for the 350~nm lattice spacing. The same analysis to obtain the average velocity at different temperatures was employed and the results are plotted in Fig.~\ref{mpvelocity}b. For the larger lattice spacing the propagation dynamics are shifted to a higher field range. Also, the velocities are in general much slower than those for the more closely coupled array. However, at the highest temperature, 511~K, similar velocities can be reached to those of the lowest temperature of the 350~nm sample. 

The principal feature of the datasets in Fig.~\ref{mpvelocity} is the regime in which velocity is linear in driving field after the onset of monopole motion at a critical field $H_\mathrm{crit}$. This can be described as 
\begin{equation}
v_\mathrm{av} = \mu_\mathrm{m} (H_\mathrm{prop} - H_\mathrm{crit}), \label{mobility}
\end{equation}
in which $\mu_\mathrm{m}$ is a monopole mobility. This is defined by analogy with electrical charge carrier mobility $\mu$ in an electric field $\cal{E}$ in semiconductor physics, where $\mu = d v / d \cal{E}$. The propagation field, $H_\mathrm{prop}$, is analogous to the electric field, providing the ``tilted'' energy landscape to drive charge drift motion. This tilt defines the preferred direction of propagation, which is the reason why the strings expand vertically in the image. Both $\mu_\mathrm{m}$ and $H_\mathrm{crit}$ were determined for each dataset in Fig.~\ref{mpvelocity} by fitting a straight line to the data in the linear regime. For the $a$~=~350~nm sample this was only done for the low field region of the data due to too many reversals within the field of view to extract a meaningful average velocity. 

\begin{figure}[tb]
\centering
\includegraphics[width=10cm]{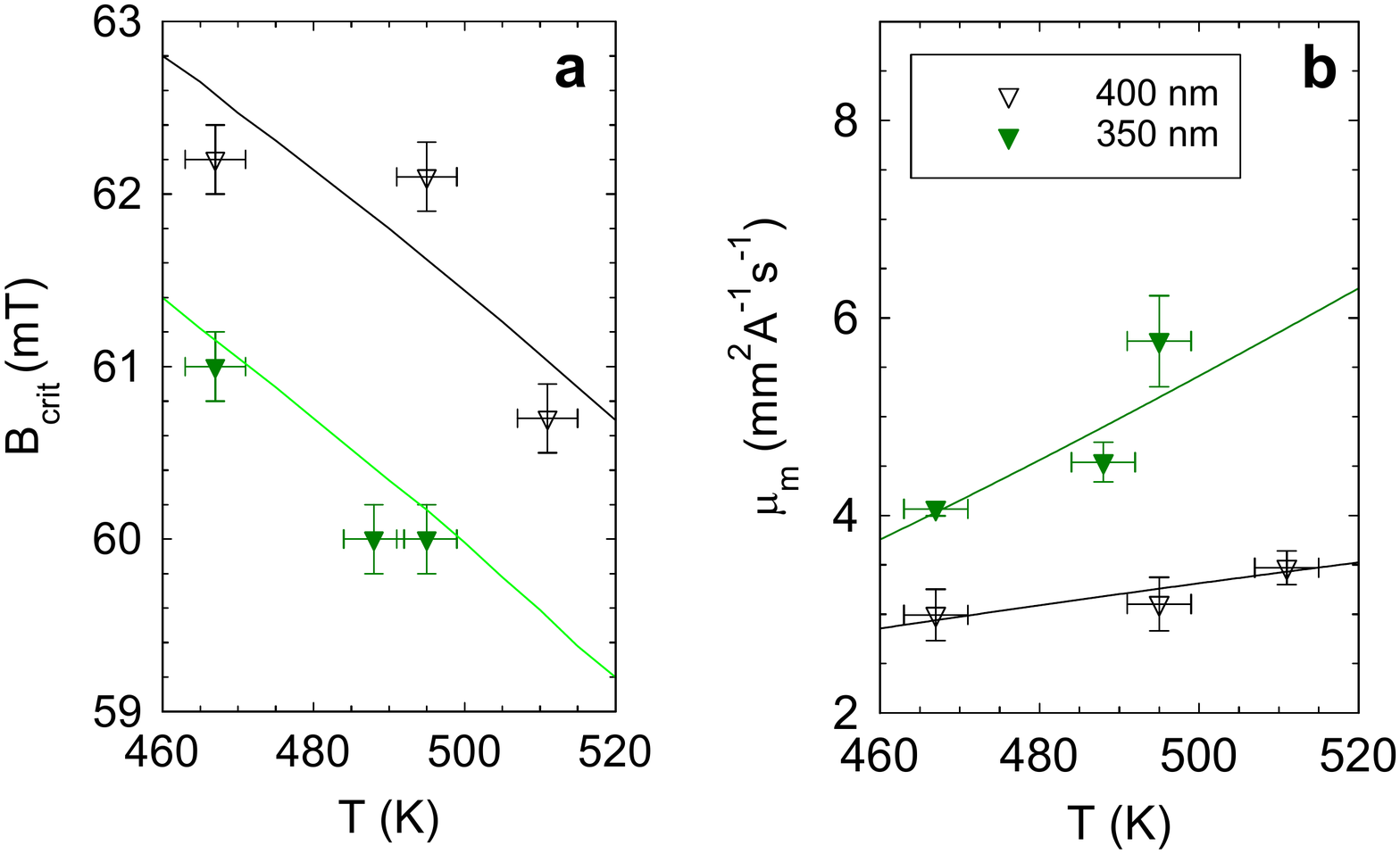}
\caption{\textsf{\textbf{Critical field and mobility for monopole drift motion.} \textbf{a}, The data points show the critical field, $B_{\mathrm{crit}} = \mu_0 H_\mathrm{crit}$, for the onset of monopole motion for arrays of two different strengths of interaction, extracted from the fits in Fig.~\ref{mpvelocity}. The solid lines are the values from the modified Bean-Livingston theory. \textbf{b}, The data points show the magnetic mobility $\mu_\mathrm{m}$ extracted from the fits in Fig.~\ref{mpvelocity}, also plotted for each array. The solid lines are fits of Eq.~\ref{lelmob} to the data. In both cases the solid symbols are for the $a = 350$~nm and the open symbols are for the $a = 400$~nm array.}}
\label{mpmobility}
\end{figure}

The results of these fits are shown in Fig.~\ref{mpmobility}, in which we display the temperature dependence of both $B_\mathrm{crit} = \mu_0 H_\mathrm{crit}$ (Fig.~\ref{mpmobility}a) and $\mu_\mathrm{m}$ (Fig.~\ref{mpmobility}b) for each of the two values of $a$. We see that $B_\mathrm{crit}$ reduces as the temperature rises and is less for a more strongly coupled array. On the other hand, $\mu_\mathrm{m}$ rises as the array is warmed, and is larger for a more strongly coupled array.
 
\subsection*{Critical field for onset of monopole motion}

In order to describe the onset of monopole motion at $H_\mathrm{crit}$, we have considered that the thermally activated reversal rate of the nanoislands is limited by the rate of hopping over the shape anisotropy energy barrier \cite{Morley2015}. For zero applied field, this barrier is given by $E_{0} = KV$, where $K = \frac{1}{2} \mu_0 \Delta N M_\mathrm{S}^2$, with $\Delta N \approx 0.1$ being the difference between the demagnetisation factors for the easy and hard axes \cite{osborn}, and $M_\mathrm{S} = 1.0 \pm 0.1$~MA/m is the measured saturation magnetisation of the Co$_{60}$Fe$_{20}$B$_{20}$. The barrier for each nanoisland must be overcome by thermal fluctuations in order to propagate the monopole excitations. The key point is that the barrier is reduced by a magnetic field \cite{Bean1959}, which in this case is a combination of the applied field $H_{\mathrm{prop}}$ and local coupling fields from neighbouring islands.

We can describe the velocity of the emergent monopoles by considering the rate of expansion of a string in both directions. In that case we have a rate $\dot{n_{t}}$ for the string to expand to the top and the same for the bottom $\dot{n_{b}}$. We can then describe the overall rate of growth of the string, $\dot{L}$, as:
\begin{equation}
\dot{L} = \dot{n_{t}} + \dot{n_{b}},
\label{dotL}
\end{equation}
where $\dot{n_{t,b}} = f_{0}(e^{- E_{>}\beta} + e^{- E_{<}\beta})$, $f_0$ is an attempt frequency, and $\beta = 1/k_{\mathrm{B}}T$. The two different energy terms here describe in one case the lengthening of the string, $E_{>}$, but also include the finite probability of a reversal against the field and a subsequent contraction of the string, $E_{<}$. We can define these using the Bean-Livingston field-adjusted barrier height \cite{Bean1959}:
\begin{eqnarray}
E_{>} & = & E_{0} \left(1 - \frac{\left[ \frac{\Delta}{m} + B_\textrm{prop} \right]}{B_{0}} \right)^{n} \\
\label{E>}
E_{<} & = & E_{0} \left(1 + \frac{\left[ \frac{\Delta}{m} + B_\textrm{prop} \right]}{B_{0}} \right)^{n}
\label{E<}
\end{eqnarray}
where $E_{0}$ is the zero field energy barrier of the system described above, $\Delta$ is the energy difference between the initial vertex type and the final vertex type $E(T_{i}) - E(T_{f})$, $m = M_{S}V$ is the magnetic moment of the island, $B_{\mathrm{prop}} = \mu_{0} H_{\mathrm{prop}}$, $B_{0}$ is the measured room temperature coercivity, and $n = 1.5$ from the geometry of the field with respect to the easy axis of the islands\cite{Victora1989}. Here the terms involving $\Delta$ are additions to the original model\cite{Bean1959} that we have used to represent the interactions between the islands.

Using eqs. \ref{dotL}~-~\ref{E<}, we were able to simulate the velocity data as a function of the propagation field. From this we could use the lowest limit of one flip in the entire measurement time as the value of field for the onset of motion, $H_{\mathrm{crit}}$, below which we would not observe any island flips beyond the injection state (\textit{i.e.} $v(H_{\mathrm{crit}})$ = 1 lattice hop per second = 0.35 or 0.40 $\mu$ ms$^{-1}$) and reproduce the temperature dependence of the measured critical fields as shown by the lines plotted in Fig. \ref{mpmobility}a. It should be noted in order to get good agreement, the temperature-corrected $M_{\mathrm{S}}$ value from Bloch's law \cite{Ashcroft} $M(T) = M(0)[1 - (T/T_{\mathrm{C}})^{3/2}]$, with $T_{C}$~=~1200~K, was used for the Co$_{60}$Fe$_{20}$B$_{20}$. In the majority of cases the strings travel in the direction of the applied field which means $\Delta$ = $E(T_{2}) - E(T_{1})$, this was estimated from numerical micromagnetic simulations, carried out using the OOMMF code\cite{oommf}, as $3.3\e{-19}$~J and $1.87\e{-19}$~J for $a$~=~350nm and 400~nm, respectively (see Methods section). We also found the hopping rate associated with $E_{<}$ to be small enough to have a negligible effect on the results. This model correctly reproduces both the temperature and coupling strength dependence of $H_\textrm{crit}$. Unfortunately, this model predicts an exponential dependence of velocity on drive field beyond $H_\textrm{crit}$, much sharper than the linear behaviour observed in the experiment. The mobility thus needs to be described using an alternative approach.

\subsection*{Monopole mobility}

In Eq. \ref{mobility} we define the monopole mobility as $\mu_\textrm{m} = dv_\textrm{av} / dH_\mathrm{prop}$. An equivalent quantity has been used used to describe field-driven domain wall motion. For instance, Beach et al. have previously defined a magnetic mobility in the study of domain wall velocity as a function of applied field in Permalloy nanowires \cite{Beach2005}. In their case they found the velocity-field curve was characterised by two regions of linearity; the low field region which has a significantly larger mobility than the high field region, in between this a negative differential mobility is observed and attributed to the Walker breakdown. That linear behaviour was in the viscous flow or precessional regimes in which thermal activation plays no role. Rather, the appropriate analogy for our thermally activated results is the so-called creep regime in which domain wall motion is frozen at 0~K but can be thermally activated at finite temperature \cite{metaxas2007}. This creep motion is usually described by a model in which a one-dimensional elastic domain wall moves through a disordered two-dimensional energy landscape that predicts a non-linear creep law in which $v_\mathrm{av} \sim \exp H^{-1/4}$, which is experimentally well-obeyed \cite{lemerle1998}. 

Nevertheless, a crossover to linear behaviour for domain wall creep velocity has been identified recently for walls in sufficiently narrow nanowires \cite{Kim2009}. This has been theoretically explained as a being due to the reduction in dimension: the domain wall becomes a zero-dimensional object moving in a one-dimensional energy landscape \cite{Leliaert2016}. That work predicts a mobility for field-driven motion that has the form 
\begin{equation}
  \mu_\textrm{m} = A \exp{\left( \frac{-\epsilon^2}{k_\textrm{B}^2T^2} \right)} \label{lelmob}
\end{equation}
where the prefactor $A = 2 \mu_0 M_\textrm{S} / \Gamma$ in which $\Gamma$ is a measure of magnetic ``friction'', $\epsilon$ is the standard deviation of the random one-dimensional energy landscape, and $k_\textrm{B}$ is the Boltzmann constant. Since our monopoles, just like domain walls, are localised magnetic excitations over a ground state, we fitted this expression to our mobility data for both values of $a$. The results are shown in Fig.~\ref{mpmobility}b. 

The prefactor $A$ is hard to determine accurately: the results of the fitting yield $A = 41 \pm 42$~mm$^2$A$^{-1}$s$^{-1}$ and $7 \pm 3$~mm$^2$A$^{-1}$s$^{-1}$ for $a = 350$~nm and 400~nm, respectively. Any difference here is not meaningful since we are trying to fit with a function which saturates to the value $A$ at large $T$ (i.e. $T \gg \epsilon$), and the data are quite far from that limit. 

The results for $\epsilon$ are more meaningful. Our fits return values of $\epsilon = (9 \pm 2) \times 10^{-21}$~J and $(6 \pm 1) \times 10^{-21}$~J for $a = 350$~nm and 400~nm, respectively. These energies correspond to temperatures $\epsilon/k_\textrm{B}$ of $600 \pm 200$~K and $400 \pm 100$~K, respectively. Stronger coupling appears to help smooth out the variations in the energy landscape, leading to higher mobility. These temperatures are comparable to the ones needed to obtain thermally activated creep in our experiments. This shows the need for thermal fluctuations on the scale of the spatial fluctuations in the one-dimensional energy landscapes to obtain linear creep motion of monopoles on laboratory timescales.

It is physically reasonable that the mobility rises sharply as the temperature approaches the scale of the typical spatial fluctuation in the energy landscape, so the temperature dependence of $\mu_\mathrm{m}$ can be understood on this basis. The fact that the mobility is higher for more strongly coupled samples can be attributed to the avalanche nature of the motion. The flipping of one island is more likely to cause the next island in the chain to flip if the coupling is stronger, leading to the monopole propagating further during that avalanche event and leading to a higher drift velocity and hence higher mobility. 

The fact that a linear dependence of $v_\textrm{av}(H_\textrm{prop})$ is observed thus means that the monopoles in the square ASI can be treated as zero dimensional point-like objects and experience a reduction in the effective dimension of their environment from two dimensions to one. A similar dimensional reduction was previously inferred for the kagom\'{e} ice on the basis of the avalanche statistics departing from a Gutenberg-Richter-type power law \cite{Mengotti2011,Huegli2012}. 

\section*{Discussion}

We have been able to image directly the thermally-activated drift motion of magnetic monopoles in artificial spin ices using a newly developed on-membrane heating device and magnetic transmission X-ray microscopy. The motion only occurs above a critical field and takes the form of one-dimensional strings that can be interpreted as Nambu strings. We have measured the drift velocity of the magnetic monopole charges as a function of drive magnetic field and temperature for arrays of different coupling strength from these images. This revealed the temperature and coupling strength dependence of both the critical field for the onset of motion and also their mobility, defined by analogy with the mobility of electric charges under an electric drive field. 

The temperature and coupling strength dependence of the critical field exhibit exponential behaviour that can be described by a Livingston-Bean model modified to include coupling fields between the elements. This model is one that is commonly used to describe the thermally activated reversal of a nanomagnet. 

The monopole drift motion is also due to thermal activation, but is found to be linear, rather than exponential, in drive magnetic field. Drawing inspiration from the crossover of domain wall creep motion from exponential to linear behaviour as the dimensionality is reduced, the equivalent dependences of the mobility have been described using a model originally developed to describe this linear form of creep motion. This allows us to extract the scale of the variations in the energy landscape through which the monopole charges propagate, which are smoothed out by stronger inter-island coupling. 

These results show that the flow of currents of `magnetricity' can be directly imaged at the individual charge carrier level in artificial spin ices and that emergent dimensional reduction for avalanches is a property of the square ice as well as the kagom\'{e} ice. 

\section*{Methods}
\subsection*{Growth and characterisation} The Co$_{60}$Fe$_{20}$B$_{20}$ alloy was first sputtered onto 100~nm-thick Si$_{3}$N$_{4}$ membranes. These were spin coated with ZEP520A: anisole (1:1) with film thickness $\approx$ 140~nm. The ASI arrays were then defined using electron beam lithography. They were written using an electron dose of 343~$\mu$C/cm$^{2}$. The pattern was developed for 70~s in N50 solution. Ti was evaporated into the pattern and lifted off to provide a hard mask for broad-beam Ar ion milling. The sample was milled for 80~s to remove the unwanted CoFeB film around the mask and leave only the CoFeB in the desired spin ice array pattern. A PMMA-based bilayer resist and another electron beam lithography step were used to create the on-membrane heater and thermometer pattern, after which Pt was evaporated at a thickness of 60~nm to create the heater and thermometer. Thermal imaging was carried out using a FLIR thermal imaging camera with a macro lens to check the temperature distribution across the membrane.

\subsection*{Soft X-ray magnetic microscopy} All the MTXM experiments were carried out at the full-field soft X-ray microscope, XM-1, located at beamline 6.1.2 at the Advanced Light Source. This microscope has a spatial resolution of about 15 nm, images can be recorded with an exposure time of a few seconds and it covers a several micrometer field of view \cite{Chao2005}. All images were taken at the Co L$_3$ edge (778 eV) with circularly polarised X-rays of a fixed helicity. This provided strong X-ray magnetic circular dichroism (XMCD) contrast arising from the high Co content of the Co$_{60}$Fe$_{20}$B$_{20}$ alloy. A nanomagnet whose moment is oriented parallel to the X-ray propagation vector will have absorption different from one that has its  anti-parallel, which provides the magnetic contrast mechanism. Since our islands are magnetised in-plane, the sample was tilted at an angle of 30$^\circ$ from normal incidence to give a magnetisation component along the beam direction. A back-thinned, back-illuminated 2048~$\times$~2048 pixel CCD camera acts as a detector to form the image, so the absorption is directly measured. A sample raw CCD image is shown in Fig.~\ref{imagingmps}a. A contrast image is obtained by dividing two consecutive absorption images, darker contrast indicates those islands that have switched their moment orientation, as shown in Fig.~\ref{imagingmps}b. The sample is aligned in such a way that the field is applied in the film plane and along a diagonal of the ASI array, so that all islands have their magnetic easy axis, which is defined by their elongated shape, at 45$^{\circ}$ to the propagation field direction.

\subsection*{Simulations} Simulations were carried out using finite element micromagnetic calculations by means of the 3-dimensional Oxsii option of the OOMMF code, in order to calculate the exchange and demagnetizing energies of the different vertex configurations. Rectangular nanoisland shapes with rounded edges were used, discretized into 2$\times$2$\times$3.5~nm$^{3}$ unit cells. The saturation magnetization $M_{\mathrm{S}}$ was determined to be 1.0$\pm$0.1~MA/m experimentally; the exchange stiffness constant $A$ used was 27.5~pJ/m, estimated from a stoichiometric average of that of Co (30~pJ/m) and Fe (20~pJ/m); and the magnetocrystalline anisotropy constant $K$ was assumed to be zero in this amorphous soft magnet. The Gilbert damping coefficient $\alpha$ was set to the unphysically high value of 0.5, allowing for rapid convergence (convergence criterion used: $dm/dt~<~$0.1~deg/ns), after obtaining similar vertex energies using 0.5 and 0.016 (where the latter simulation time is much longer). 

\bibliography{MorleyThesis_bib}

\section*{Acknowledgements}

We would like to thank Gianfranco Durin for helpful discussions. We acknowledge support from EPSRC Grants EP/L00285X/1 and EP/L003090/1. The Advanced Light Source is supported by the Director, Office of Science, Office of Basic Energy Sciences, of the U.S. Department of Energy under Contract No. DE-AC02-05CH11231. Mi-Young Im acknowledges support by Leading Foreign Research Institute Recruitment and Future Materials Discovery Programs through the National Research Foundation (NRF) of Korea funded by the Ministry of Education, Science and ICT (2018K1A4A3A03075584, 2016M3D1A1027831, 2017R1A4A1015323) and by the DGIST R\&D program of the Ministry of Science, ICT and Future Planning (18-BT-02). We would like to acknowledge Dr Mike Cooke for use of the infra-red microscope at the University of Durham.

\section*{Author contributions statement}

Sample design and preparation: S.A.M. and M.C.R.; measurements: S.A.M., J.M.P., A.H., M-Y.I., P.J.F., S.L. and C.H.M.; simulations: J.M.P.; analysis and interpretation: S.A.M., J.M.P., G.B., C.H.M.; supervision of the project:  S.L. and C.H.M.. E.H.L. provided support for M.C.R.. All authors contributed to and reviewed the manuscript.

\section*{Additional information}

\noindent \textbf{Competing financial interests:} The authors declare no competing financial interests.

\noindent \textbf{Reprints and permissions} information is available online at http://npg.nature.com/reprintsandpermissions.

\noindent Correspondence and requests for materials should be addressed to S.A.M. (samorley@ucsc.edu) or C.H.M. (c.h.marrows@leeds.ac.uk).

\end{document}